\documentstyle[epsfig, twocolumn]{iso98}

\newcommand{\thepapername}{}

\def\etal{{\em et al.}\ }
\def\deg{\hbox{$^\circ$}}
\def\arcmin{\hbox{$^\prime$}}
\def\arcsec{\hbox{$^{\prime\prime}$}}
\def\micron{\hbox{$\mu$m\ }}
\begin{document}

\setlength{\parindent}{0pt}
\setlength{\parskip}{ 10pt plus 1pt minus 1pt}
\setlength{\hoffset}{-1.5truecm}
\setlength{\textwidth}{ 17.1truecm }
\setlength{\columnsep}{1truecm }
\setlength{\columnseprule}{0pt}
\setlength{\headheight}{12pt}
\setlength{\headsep}{20pt}
\pagestyle{veniceheadings}

\title{\bf THE EUROPEAN LARGE AREA ISO SURVEY: ELAIS}

\author{
{\bf M.Rowan-Robinson$^{1}$, 
S.Oliver$^{1}$,
A. Efstathiou$^{1}$,
C. Gruppioni$^{1}$,
S. Serjeant$^{1}$,}\\
{\bf C. Cesarsky,
L. Danese,
A. Franceschini,
R. Genzel,
A. Lawrence,
D. Lemke,}\\
{\bf R. McMahon,
G. Miley,
I.Perez-Fournon,
J-L. Puget,
B. Rocca-Volmerange,}\\
{\bf P. Ciliegi,
P. H\'eraudeau, 
C. Surace,
F.LaFranca
and ELAIS consortium}
 \vspace{2mm}\\
$^1$Imperial College of Science Technology and Medicine,\\
	   Astrophysics Group,
	   Blackett Laboratory,\\
	   Prince Consort Rd., London,
	   SW2 1BZ,\\ 
	   mrr@ic.ac.uk}

\maketitle

\begin{abstract}

  The European Large Area {\em ISO} Survey (ELAIS) has surveyed  $\sim 12$ square
degrees of the sky at 15\micron and 90\micron and subsets of this
area at 6.75\micron and 175\micron using the  Infrared Space Observatory 
({\em ISO}).  This project was the largest single open time programme executed
by {\em ISO}, taking 375 hours of data.
A preliminary catalogue of more than 1000 galaxies has been
produced.   In this paper we describe the goals of the project,
describe the follow-up programmes that are in progress,  
and present some first scientific results including
a provisional number count analysis at 15 and 90\micron.
  \vspace {5pt} \\


  Key~words: ISO; infrared astronomy; surveys; starburst galaxies; AGN.

\end{abstract}

\section{INTRODUCTION}

The
European Large Area {\em ISO} Survey (ELAIS) is 
a collaboration involving 19 European institutes (in
addition to the authors and others at their institutes
the following people and others at their institutes are involved
I. Gonzalez-Serrano,
E. Kontizas,
R. Mandolesi,
J. Masegosa, 
K. Mattila,
H. Norgaard-Nielsen,
M. Ward)
and is the
largest open time project being undertaken by {\em ISO}.

In this paper we describe some of the key scientific goals of ELAIS and
give a brief description of the survey, comparing it with other ISO 
surveys.  We give a preliminary discussion of the  15 and 90 $\mu$m
source-counts and describe the follow-up programme now under way at
a wide range of wavelengths, giving some first scientific results.

\section{KEY SCIENTIFIC GOALS}

In this section we highlight some of the principal 
scientific motivations for this programme, though naturally there
are many other goals which we do not have space to discuss.

\begin{itemize}
\item The main extra-galactic population detected by IRAS was galaxies with
high rates of star formation. These objects are now known
to evolve with a strength comparable to AGN. 
 The sensitivity of {\em ISO} 
allow us to detect these objects at much higher redshifts and thus
obtain greater understanding of the cosmological evolution of star
formation.  This is directly complementary to studies of
star-formation history in the optical and UV. 
Comparison of the star-formation rates determined in the FIR
with that determined from the UV will give a direct estimate of the
importance of dust obscuration, vitally important for models of
cosmic evolution.


\item If elliptical galaxies underwent a massive burst of star-formation
between $2<z<5$, they would be observable in the far infrared and may
look like IRAS F10214+4724.  This survey will provide a powerful
discrimination between such a top-down scenario and any hierarchical
bottom-up merging model whose components are
individually too faint to detect.


\item Unified models of AGN suggest that the central engine is surrounded by
a dusty torus.  The mid and far infrared emission from the torus  is much
less sensitive to the viewing angle than the optical.  Thus a mid and far
infrared selected sample of AGN will place
important constraints on unification schemes.


\item IRAS uncovered a population with enormous far infrared luminosities,
$L_{\rm FIR} $ $> 10^{12}L\odot$.    While many of
these objects appear to contain an AGN it is argued that star formation
could provide most of the energy.  Interestingly, most of these
objects appear to be in interacting systems, suggesting a
triggering mechanism.  Exploration of this population to higher redshift
will have particular significance for models of AGN/galaxy evolution.
\end{itemize} 

\section{THE ELAIS SURVEY}

The {\em ISO} observations began on 12th March 1996 and ended on 8th April 1998,
shortly before the exhaustion of the Helium coolant.  The total survey area covered in
blank fields is 6, 11, 12, \& 1 sq. deg. at 6.75, 15, 90 and 175 \micron,
and an additional 2 sq. deg. within the ELAIS survey areas has been surveyed
at 175 $\mu$m by the FIRBACK team (Puget et al 1999).  Table 1 specifies the main
survey areas.  A full description of the survey will be given by Oliver et al (1999).

As part of the follow-up programme described below, we have embarked on surveys of
these same areas at X-ray, ultraviolet, visible, near infrared, submillimetre and radio
wavelengths. Table 2 shows the wavelengths and areas covered (or planned to be covered
in allocated observing time) across the whole electromagnetic spectrum.  The surveys
at 2-10 keV, U, K, 850 $\mu$m and 21 cm will be among the largest area surveys in these
wavebands, to these sensitivities, being undertaken.  Thus ELAIS has developed from
its original concept as a survey at 15 and 90 $\mu$m into a powerful multiwavelength
survey across the entire electromagnetic spectrum.  

\begin{center}
\begin{table*}
\caption{Summary of ELAIS Survey Areas.
These areas were selected primarily for having low
Cirrus contamination, specifically $I_{100}<1.5$MJy/sr from
the IRAS maps of Rowan-Robinson \etal (1991).
For N1-3, S1 and all X areas we also restricted ourselves to regions of
high visibility $>25$\% over the mission lifetime.  For low Zodiacal 
background we
required  $|\beta|>40$ and to avoid saturation of
the CAM detectors we had to avoid any bright IRAS 12$\mu$m sources.
6 additional small ($0.3\deg \times 0.3\deg$) rasters (X1-6)
were centred on well studied areas of the
sky or high-$z$ objects}
\label{areas}
\begin{tabular}{lrrrrrc}
&&&&&&\\
Name &
\multicolumn{2}{c}{Nominal Coordinates} &
M & N & ROLL &
 $\langle I_{100}\rangle$ \\

  & \multicolumn{2}{c}{J2000}
 & /\deg & /\deg & /\deg
 & \begin{small}$/{\rm MJy sr}^{-1}$ \end{small} \\
\hline
\\
N1&$16^h10^m01^s$&$+54\deg30\arcmin36\arcsec$&
 2.0 & 1.3 & 20 & 1.2 \\
N2 &$16^h36^m58^s$&$+41\deg15\arcmin43\arcsec$&
 2.0 & 1.3 & 30 &  1.1  \\
N3 &$14^h29^m06^s$&$+33\deg06\arcmin00\arcsec$&
 2.0 & 1.3 & 330 &  0.9 \\
S1 &$00^h34^m44^s$&$-43\deg28\arcmin12\arcsec$&
 2.0 & 2.0 & 20 &  1.1 \\
S2 &$05^h02^m24^s$&$-30\deg35\arcmin55\arcsec$&
0.3 & 0.3 & 290 & 1.1 \\
\hline
\end{tabular}
\end{table*}
\end{center}

\begin{center}
\begin{table*}
\caption{Field Surveys within the ELAIS regions, the vast majority carried out as part 
of the ELAIS collaboration.}

\begin{small}

\begin{tabular}{lccccccccccc}
&&&&&&&&&&&\\
Band       & 2-10keV &  $U$  & $B,V,I$ & $R$   & $H/K$ & 6.7 & 15  & 90 & 175  &
 850  & 21cm  \\
\hline
\\
Area (sq degs)       & 0.15     &  6    &    1    & 13    &   1   &  6  & 11  & 12 &  3   &
 0.1 &  8.2  \\
Depth      & $1\times 10^{-14}$
                     & 22.5  &   23    & 23.5  & 19.5  & 1   & 3   & 100 & 100 &
 8 &  0.1-0.4 \\
Units      & CGI     &  mag  &   mag   & mag   & mag   & mJy & mJy & mJy & mJy &
 mJy  & mJy   \\
Galaxies   &    &  30000  &         &       & 3000  & 1104&1618 & 390 & 100 &
      & 1448  \\
Stars      &         &  15000  &         &       &       &     &     &     &     &
      &       \\
\hline
\end{tabular}
\end{small}
\end{table*}
\end{center}

\section{COMPARISON WITH OTHER {\em ISO} SURVEYS}

 {\em ISO} carried out a variety of surveys exploring the available 
parameter space of depth and area (cf the review by Oliver (1998)).  Table 3 
summarizes
the main ISO extra-galactic blank-field surveys.  It is clear that information
will be available at a wide range of ISO wavelengths, and to a wide range of
sensitivities, and that together these surveys will provide a powerful probe of
the infrared sky.  The ELAIS survey will play a key role because of its large area
and the numbers of sources detected (see Table 2 for numbers found in our Quick-Look
Analysis).

\begin{table*}
\begin{center}
\caption{Field Surveys with {\em ISO} \label{surveys}}
\begin{tabular}{lllll}
&&&&\\
Survey Name  & [e.g. ref] & Wavelength & Integration & Area\\
             &    &   $/\mu$m  &   $/$s      & $/{\rm sq deg}$\\
\hline
\\
PHT Serendipity Survey & Bogun et al 96     & 175         & 0.5            & 7000 \\
CAM Parallel Mode & Siebenmorgen et al 96       & 7           & 150            & 33 \\
ELAIS         &     & 7,15,90,175 & 40, 40, 24, 128& 6, 11, 12,1\\
CAM Shallow   &   Elbaz et al 98         & 15          & 180            & 1.3  \\
FIR Back      &    Puget et al 99 & 175         & 256, 128       & 1, 3  \\
IR Back       &    Mattila et al 99         & 90, 135,180 & 23, 27, 27     & 1, 1, 1 \\
SA 57         & Norgaard-Niielson et al 97 & 60, 90      & 150, 50        & 0.42,0.42  \\
CAM Deep      &    Elbaz et al 98    & 7, 15, 90   & 800, 990, 144  & 0.28, 0.28, 0.28\\
Comet fields   &   Clements et al 99 & 12          & 302            & 0.11\\
CFRS          & Hammer and Flores 98                    & 7,15,60,90  & 720, 1000, 3000,3000 & 0.067\\
CAM Ultra-Deep&       Elbaz et al 98         & 7           & 3520           & 0.013 \\
ISOHDF South  & Oliver et al 99b           & 7, 15       &$>6400, >6400$  & 4.7e-3, 4.7e-3 \\
Deep SSA13    & Taniguchi et al 97                & 7           & 34000          & 2.5e-3\\
Deep Lockman  &  Kawara et al 98     & 7, 90, 175  & 44640, 48, 128 & 2.5e-3, 1.2 , 1 \\
ISOHDF North  & Serjeant et al 97        & 7, 15       & 12800, 6400    & 1.4e-3, 4.2e-3 \\
\hline
\end{tabular}
\end{center}
\end{table*}

\section{PRELIMINARY NUMBER COUNTS}

Source catalogues have been extracted from the ELAIS data at all 
wavelengths and a preliminary source count analyses have been
performed at 15 and 90 \micron.  The 15 $\mu$m results are shown in Figure
1, where we have included results from our HDF-N survey (Oliver et
al 1997, Aussel et al 1998), from the ISO Deep survey in the Lockman hole
(ref) and extrapolated from IRAS 12 $\mu$m counts (Verma 1999).  The models
shown are from Pearson and Rowan-Robinson (1996) and agree well with the data.
Similar agreement is found for the models of Franceschini et al 
(1997).   As can been 
seen these counts confirm the strong
evolution detected in the {\em ISO HDF} analysis (Oliver et al 1997).
A more detailed analysis of the 15 $\mu$m counts will be presented shortly 
(Serjeant et al 1999).  

The counts at 90 $\mu$m (Fig 2) are extremely preliminary, since there are still
major uncertainties in the calibration and there is so far no detailed
correction for incompleteness due to the variable effects
of cosmic rays.  However they show reasonable consistency at the bright end
with counts at 90 $\mu$m interpolated from IRAS 60 and 100 $\mu$m data
and demonstrate that we should be reasonably complete
at least to 100 mJy.  A more detailed discussion of the 90 $\mu$m counts will
be given by Efstathiou et al (1999).

Results from our northern and southern 21 cm surveys have already been published 
(Ciliegi et al 1999, Gruppioni et al 1999) and show good consistency with
earlier sub-mJy radio surveys.  These radio surveys represent a very substantial
expansion of the area of the sky surveyed to sub-mJy sensitivities and the follow-up
of these radio surveys will be a very interesting project in its own right.

%
\begin{figure}
\centerline{\vbox{
\psfig{figure=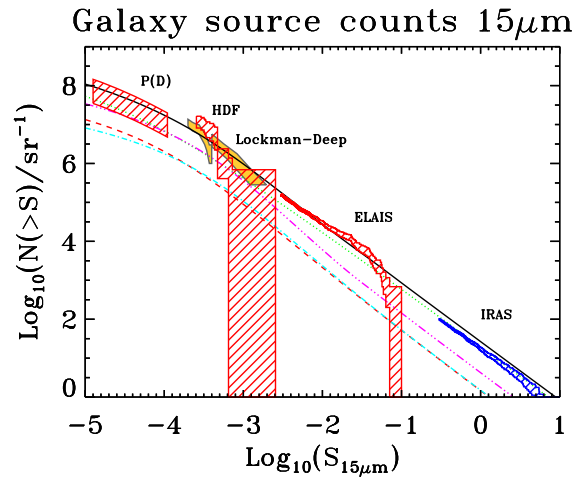,height=9.cm,angle=90}
}}
\caption[]{Preliminary 15 \micron integral source 
counts from the ELAIS survey (Serjeant et al 1999),
together with counts from {\em ISO HDF} (Oliver et al 1997, hatched, 
Aussel et al 1998, shaded), and CAM Deep survey (Elbaz et al 1998, shaded),
with {\em IRAS} 12 $\mu$m counts (hatched) at bright end, 
( {\em IRAS} data shifted to 15 $\mu$m using 
cirrus spectrum (Verma 1999)).  Models are
from Pearson and Rowan-Robinson (1996): 
all components solid;
spiral galaxies dotted;
star-bursts dash-dot-dot-dot;
AGN-dash.
Faint end constraints (hatched) come from the {\em ISO HDF} $P(D)$ analysis.}
\label{af_15}
\end{figure}

\begin{figure}
\centerline{\vbox{
\psfig{figure=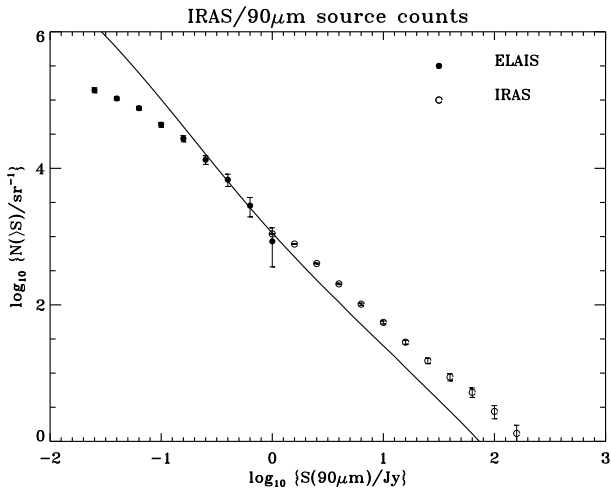,height=9.cm,angle=90}
}}
\caption[]{Preliminary 90 \micron integral source 
counts from the ELAIS survey (filled circles),
together with counts interpolated from {\em IRAS} 60 and 100 $\mu$m data (open circles) 
(Verma 1999).  Model is
from  Rowan-Robinson (1999).}

\end{figure}

\begin{figure}
\centerline{\vbox{
\psfig{figure=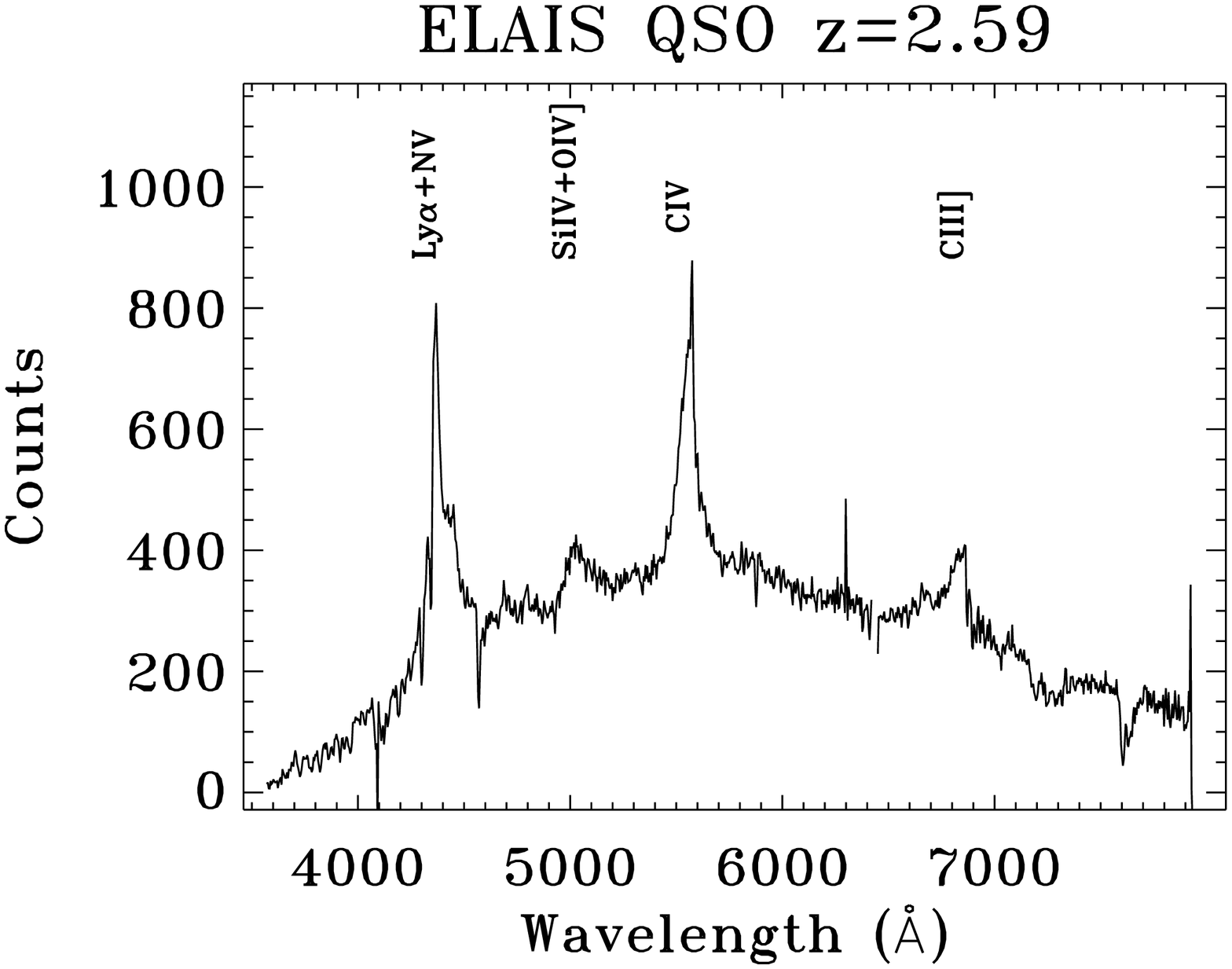,height=9.cm,angle=90}
}}
\caption[]{Spectrum of a z=2.59 QSO associated with an ELAIS 15 $\mu$m source, taken
with the 2dF on the AAT (Gruppioni et al 1999).}

\end{figure}

\section{FOLLOW-UP}

An extensive follow-up programme is being undertaken,
including measurements at all wavelengths from X-ray to radio (Tables 2 and 4).  This
programme will provide essential information for identifying the
types of objects detected in the infrared, their luminosities,
energy budgets and other detailed properties.  As well as
studying the properties of the objects detected by {\em ISO} a number
of the follow-up surveys will provide independent source lists
which will be extremely valuable in their own right, not least because
we can investigate why some objects emit in the infrared and others
do not.  As we emphasized above ELAIS has become a deep multi-wavelength
survey of these 12 sq degrees of sky, which will have impacts on a wide
range of extragalactic astrophysics.  The wealth of multi-wavelength information
will make ELAIS areas a natural focus for future surveys, for example with
WIRE, SIRTF, XMM and FIRST.

Preliminary optical identifications have already been made for several hundred
15 and 90 $\mu$m and 21 cm sources and these have been the subject of several
spectroscopic runs (see Table 4).  Results from spectroscopy of 90 $\mu$m
galaxies with R $<$ 18 in the south using FLAIR and the ESO/Danish 1.5m telescope will be
published shortly (Linden et al 1999).  As expected the galaxies tend to be
normal spirals or starbursts.  We will also be reporting the first results from
spectroscopy of 15 $\mu$m and 21 cm galaxies using 2dF and the ESO 3.6m and NTT
telescopes (Gruppioni et al 1999, La Franca et al 1999).  Figure 3 shows the spectrum
of a z=2.59 quasar found in our first 2dF run.  Taking into account those 
galaxies which are too faint for their spectra to be classifiable, about 20-40 $\%$ of the 
15 $\mu$m galaxies for which we have spectra so far
are AGN (QSO, Sey 1, Sey 2 or NELG), broadly consistent with
expectations of the models of Pearson and Rowan-Robinson (1996) and 
Franceschini et al (1998).

In conclusion it is clear that the ELAIS project will not only meet its original
scientific goals, but will also provide the basis for a series of powerful
multiwavelength surveys over the next decade.

\begin{table*}
\begin{center}
\caption{ELAIS Follow-up Programme Summary}
\begin{tabular}{lll}
&&\\
Telescope & shifts$/$nights & Projects \\
\hline
\\
AXAF & 150ks & X-ray Survey in N1, N2 \\
ESO 3.6m & 3 & Optical Imaging \\
 & & Optical Sepctroscopy \\
ESO NTT & 9 & NIR Photometry \\
 & & Optical Spectroscopy \\
ESO 2.2m & 13 & NIR Photometry \\
ESO/Dan. 1.5m & 35 & Optical Survey \\
 & & Optical Photometry \\
 & & Optical Spectroscopy \\
AAT & 2 & Optical Spectroscopy (2dF) \\
 & & NIR Spectroscopy \\
INT & 13 & Optical Survey \\
 & & NIR Survey \\
CFHT & 1 & Integral Field Spectroscopy \\
Calar Alto 3.6m & 8 & Optical Spectroscopy \\
 & & NIR Photometry \\
Calar Alto 2.2m & 2 & Optical Spectroscopy \\
UK Schmidt & 2 & Optical Spectroscopy (FLAIR) \\
Mt.Hopkins & 12 & NIR Photometry \\
ITRF & 5 & NIR Photometry \\
IAC 0.8m & 28 & Optical Photometry \\
ESA OGS 1.m & 4 & Optical photometry\\
ATNF CA & 10 & 21 cm Survey \\
VLA & 6 & 21 cm Survey \\
\hline
\end{tabular}
\end{center}
\end{table*}

\section{ACKNOWLEDGEMENTS}

This paper is based on observations with {\em ISO}, an ESA project, with
instruments funded by ESA Member States (especially the PI countries:
France, Germany, the Netherlands and the United Kingdom) and with
participation of ISAS and NASA.
This work was in part supported by PPARC grant no.  GR/K98728
and  EC Network is FMRX-CT96-0068.
 


\end{document}